\pdfoutput=1
\documentclass[aps,prl,twocolumn,showpacs,preprintnumbers,amsmath,amssymb,groupedaddress]{revtex4}

\usepackage{graphicx}
\usepackage{dcolumn}
\usepackage{bm}

\begin{document}

\title{Shot Noise and Fractional Charge at the 2/3 Composite Fractional Edge Channel}

\author{Aveek Bid}
\affiliation{Braun Center for Submicron Research, Department of Condensed Matter Physics,
Weizmann Institute of Science, Rehovot 76100, Israel}
\author{N. Ofek}
\affiliation{Braun Center for Submicron Research, Department of Condensed Matter Physics,
Weizmann Institute of Science, Rehovot 76100, Israel}
\author{M. Heiblum}
\email[e-mail: ]{moty.heiblum@weizmann.ac.il}
\affiliation{Braun Center for Submicron Research, Department of Condensed Matter Physics,
Weizmann Institute of Science, Rehovot 76100, Israel}
\author{V. Umansky}
\affiliation{Braun Center for Submicron Research, Department of Condensed Matter Physics,
Weizmann Institute of Science, Rehovot 76100, Israel}
\author{D. Mahalu}
\affiliation{Braun Center for Submicron Research, Department of Condensed Matter Physics,
Weizmann Institute of Science, Rehovot 76100, Israel}

\begin{abstract}
The exact structure of edge modes in `hole conjugate' fractional quantum Hall states remains an unsolved issue despite significant experimental and theoretical efforts devoted to their understanding.  Recently, there has been a surge of interest in such studies led by the search for neutral modes, which in some cases may lead to exotic statistical properties of the excitations.  In this letter we report on detailed measurements of shot noise, produced by partitioning of the more familiar 2/3 state.  We find a fractional charge of (2/3)\emph{e} at the lowest temperature, decreasing to \emph{e}/3 at an elevated temperature.  Surprisingly, strong shot noise had been measured on a clear 1/3 plateau upon partitioning the 2/3 state.  This behavior suggests an uncommon picture of the composite edge channels quite different from the accepted one.
\end{abstract}



\maketitle



The fractional quantum Hall (FQH) effect ~\cite{one, two} is a phenomenon observed in a high-mobility two dimensional electronic gas (2DEG) subjected to perpendicularly applied quantizing magnetic field.  The transverse magnetoresistance exhibits plateaux as function of magnetic field or electron density at rational fractions $\nu$ of the quantum conductance $e^2/h$, with simultaneous vanishing longitudinal resistance.  This phenomenon is a direct consequence of the gapped many body ground state formed at the specific filling factors $\nu$ ~\cite{three}. The lowest energy excitations of these fractional states (quasiparticles) carry fractional electronic charges ~\cite{three} and are predicted to have fractional statistics ~\cite{four}.  Wen showed ~\cite{five} that the transport should resemble that in the integer quantum Hall effect (IQHE), namely, via edge channels (each is a chiral one dimensional Luttinger liquid). Subequently, it was theoretically proposed ~\cite{six}, and later experimentally verified ~\cite{seven} that shot noise measurements can be used to probe these fractionally charged quasiparticles.  Although the FQH states are a manifestation of strong electron correlations, the quasiparticles are only weakly interacting.  Thus, the \emph{zero frequency} spectral density of the shot noise (introduced via partitioning) can be well accounted for by an analytic expression that is strictly valid for non-interacting particles  ~\cite{eight, nine, ten}.

Transport is more complicated if there are one or more counter-propagating edge channels, as is the case of the so called hole conjugate quantum hall states, $\nu$ =2/3, 3/5, etc.  In the case of $\nu$ =2/3, which is the subject of our present work, a clean sample devoid of any impurities is expected to support two charged modes: one with conductance of $e^2/h$ - carrying electrons, and a counter-propagating one with conductance (1/3)$e^2/h$, carrying $e$/3 fractional charges ~\cite{eleven, twelve, thirteen}.  Inclusion of interaction between the channels leads to a non-universal Hall conductance (in contrast to experimental observations), which depends on the interaction strength ~\cite{fourteen}.  For a smooth edge potential, the two counter-propagating modes will have different momenta (the difference being proportional to the enclosed flux), and hence unlikely to equilibrate.  However, in the presence of random inter-channel scattering the momentum of each channel need not be conserved, allowing thus equilibration (and the emergence of a single charge mode) and a universal quantization of Hall conductance (2/3)$e^2/h$.  In addition, a neutral counter propagating channel (namely, carrying only energy but no charge) is expected to exist ~\cite{thirteen, fourteen}.  Although the charge mode had been detected using time resolved transport measurements ~\cite{fifteen}, experimental evidence of the neutral mode is still missing.  Recently, there has been a resurgence of theoretical investigations of this and similar neutral edge channels ~\cite{sixteen, seventeen, eighteen}; ignited by the hypothesis of the much anticipated non-abelian $\nu$ =5/2 fractional state, which is expected to carry a neutral Majorana mode ~\cite{seventeen, eighteen}.  Here, we concentrate on the 2/3 fractional state - being the `simplest' fraction that is postulated to have a neutral mode.

The quantum shot noise is a direct result of partitioning (backscattering), routinely achieved by tunneling bringing to a close proximity the two counter propagating edge channels by a narrow constriction in the 2DEG.  The tunnelling operator of the quasiparticles is related to a local scaling dimension of the tunnelling quasiparticle $\Delta$  via $e^{-\Delta}$.  Using the model of charged and neutral modes for $\nu$ =2/3, it was predicted that quasiparticles with charge $e$/3, 2$e$/3 and $e$ should all have the same scaling dimension ~\cite{thirteen}, and consequently only the details of the structure determine the tunneling charge.  Hence, in our experiments we measured a few different structures fabricated on two separately grown 2DEG.

The narrow constriction in the 2DEG is provided by a quantum point contact (QPC) - a lithographically patterned split metallic gate deposited on top of a Hall-bar mesa.  The application of a negative voltage to the split gate with respect to the 2DEG forms a narrow constriction in the 2DEG.  While the channels that pass freely through the constriction do not carry noise at zero temperature ~\cite{nineteen}, the partitioned channels carry shot noise.  Similar to classical shot noise, the `low frequency' spectral density is proportional to the DC excitation current and to the charge of the quasiparticles.  In multiple channel transport only the partitioned channel carries noise, which is independent of the presence of other channels that are fully transmitted or are fully reflected.  Indeed, in previous measurements, the assumption of mutually independent propagating modes was found to strictly hold ~\cite{twenty, twentyone, twentytwo}.

Two GaAs-AlGaAs heterostructures, with embedded high mobility 2DEG, were used.  One (labeled A) had a low temperature mobility in excess of 6 $\times  10^6$ cm$^2$/V~s and an electron density 8.8$\time 10^{10}$cm$^{-2}$, and another (labeled B) had a mobility 4.3$\times  10^6$ cm$^2$/V~s and an electron density 10$\time 10^{10}$cm$^{-2}$.  Four different structures had been fabricated (different processes and different QPC configurations, with three of them on the higher mobility 2DEG).  The QPCs were made either by top metallic split-gate or via `mesa-side-gates' ~\cite{twentytwo} - these two methods provide very different confining potentials for the constrictions.  The data taken in all samples was found to be quantitatively very similar.  The measurements, unless specifically mentioned, were carried out in a dilution refrigerator at an electron temperature of 10mK (as deduced from shot noise measurements).
\begin{figure}[b]
\includegraphics[viewport=1in 3.3in 8in 8in,width=0.9\columnwidth]{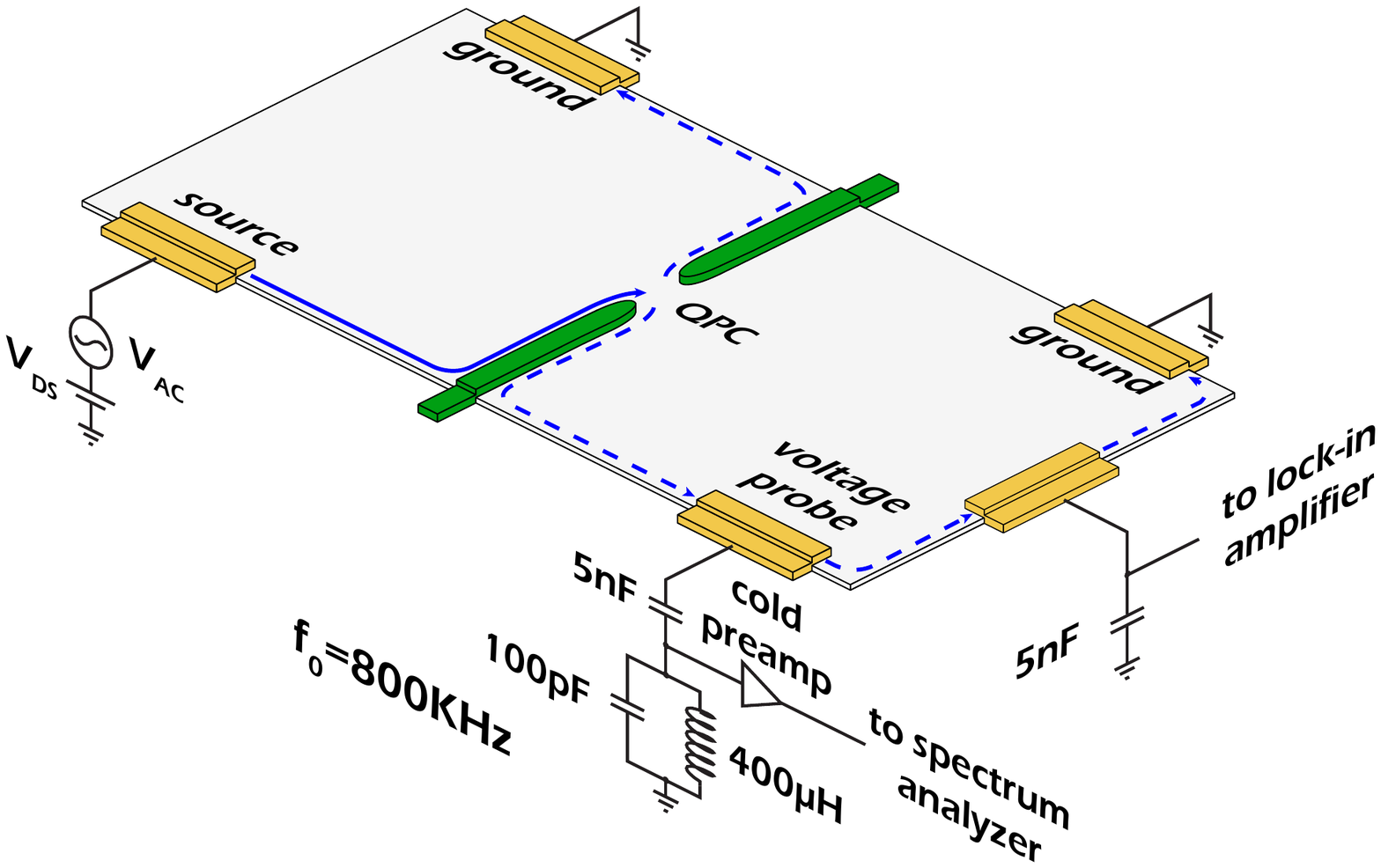}
\caption{Schematic of the noise measurement setup.  (See text for details.)}
\label{fig:figure1}
\end{figure}

The configuration of the device is shown in Fig.~1.  A split-gate, with 400nm gap was deposited on the surface of the heterojunction, forming upon biasing a controlled constriction in the 2DEG.  The multi-terminal configuration ensures a constant output resistance at the drain at a Hall plateau (being Hall resistance) - independent of the transmission of the constriction, thus allowing subtracting the contribution of the `current noise' of the preamplifier ~\cite{twentyfour}.  The fluctuations in the drain voltage were $I_dR_q$, with $I_d$ the current fluctuations and $R_q$ the quantum resistance for bulk filling factor $\nu$.  The drain voltage was filtered by a resonant circuit tuned to $\sim$800KHz with a bandwidth of some 30kHz, and subsequently amplified by a homemade, low-noise, cryogenic preamplifier (cooled to 4.2K, with voltage noise $\sim$800pV Hz$^{-1/2}$ and current noise $\sim$10fA Hz$^{-1/2}$).  The output of this preamplifier was fed to a room temperature amplifier followed by a spectrum analyzer.  Note that the central frequency was chosen to be far above the 1/\emph{f} noise knee of the sample, with the 1/\textit{f} noise contribution (which is quadratic with the current) much smaller than the shot noise and the thermal noise.  All measurements in the fractional regime were preceded by charge measurements in the integer regime, verifying that an electron charge is being measured.
\begin{figure}[b]
\includegraphics[viewport=3.5in 0in 8in 5.9in,width=0.8\columnwidth]{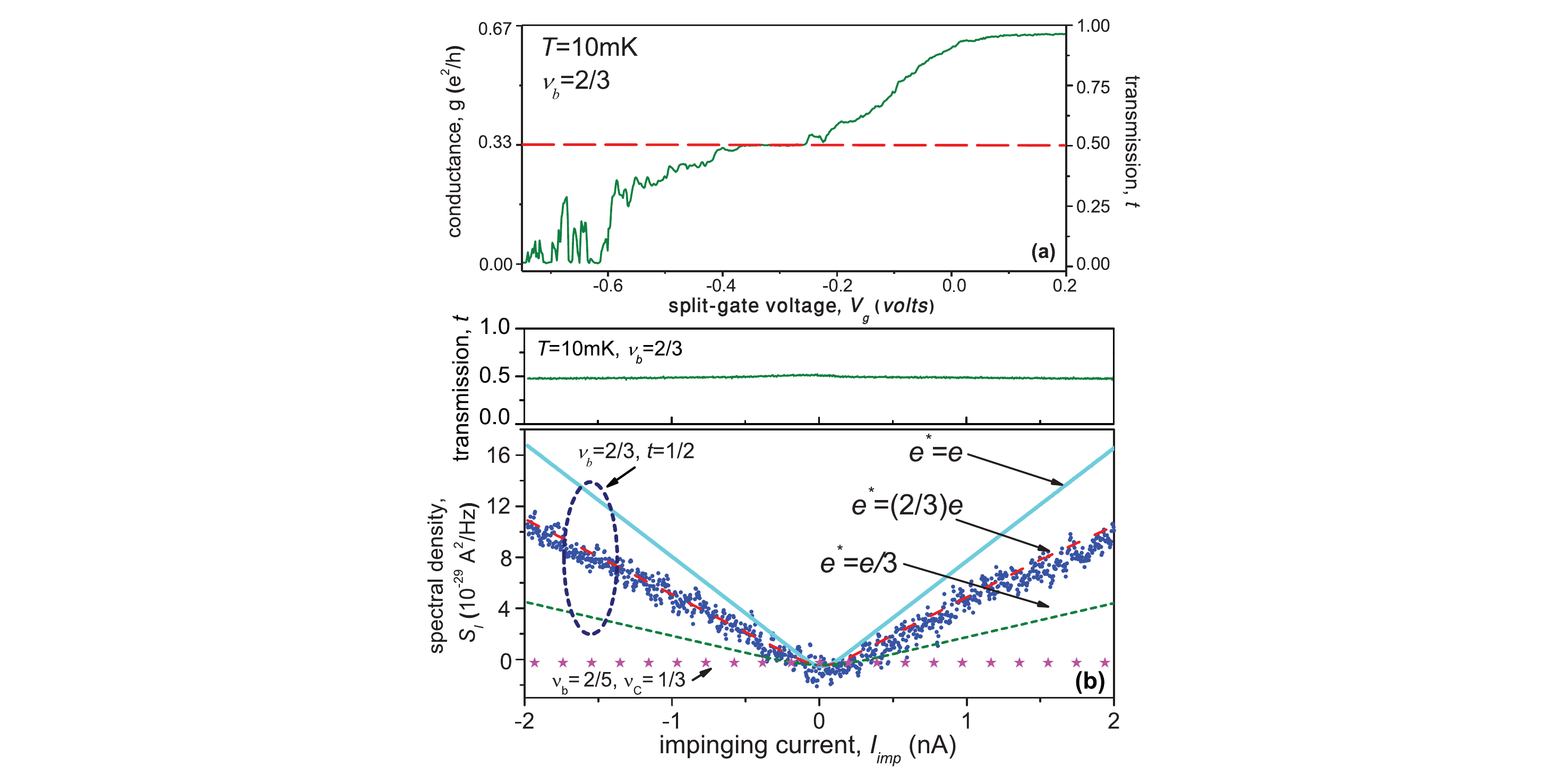}
\caption{Conductance and spectral density at electron temperature 10mK.  (a) Conductance \emph{g} and transmission \emph{t} of the constriction as a function of split-gate voltage.  Note the appearance of a prominent plateau at \emph{g}=$e^2$/3$h$ ($t$=1/2).  (b) Upper panel - dependence of the transmission (zero bias \emph{t}=1/2, split-gate voltage $V_g$ = - 0.3 V) on injected electron energy.  Lower panel - spectral density $S_I$ at this value of transmission.  The blue dots are the measured data points.  Shown is the expected spectral density for transmission \emph{t}=1/2, temperature \emph{T} = 10mK and quasiparticle charge $e^* = e$ (solid cyan line), (2/3)e (dashed red line) and e/3 (dotted olive line).  For comparison, we also show (purple stars) the noise measured when  $\nu_b$ = 2/5 and $\nu_c$ = 1/3. }
\label{fig:figure2}
\end{figure}

The spectral density of a partitioned current due to stochastic back scattering at a finite temperature is described well by the analytic expression ~\cite{eight, nine, ten}:
\begin{equation} \nonumber
S_I(0)=2eI_{imp}t(1-t)[coth(e^*V/2kBT)-2kBT/e^*V],
\end{equation}
where the impinging current $I_{imp}=Vg_q$ with $g_q$=(2/3)$e^2/h$ for bulk filling factor $\nu$=2/3, \textit{t} the constriction's transmission coefficient (assuming energy independence), $e^*$ the quasiparticle charge, and \emph{T} the electron temperature.  When \textit{t} depends weakly on the current, its differential value as function of current was used.  Figure~\ref{fig:figure2}a  shows a plot of the transmission, deduced from the two terminal linear conductance \textit{g}, as function of the applied split-gate voltage to the QPC.  At zero gate voltage, or even at slightly positive voltage, the transmission was less than unity (by a few percent); possibly due to a marginally induced potential below the split-gate.  Scanning the gate voltage revealed a prominent plateau at \textit{t}=1/2, suggesting a local filing factor in the constriction $\nu_c$=1/3.  Adopting the assumption that the electron density drops gradually near edges (within the constriction), the $\nu_c$=1/3 plateau seems to confirm that 1/3 edge channel traverses the constriction without backscattering while the 2/3 edge channel is fully reflected. Under such circumstances the shot noise for \emph{t}=1/2 should be zero. 
\begin{figure}[b]
\includegraphics[viewport=0.5in 1.2in 3.6in 4.5in,width=0.75\columnwidth]{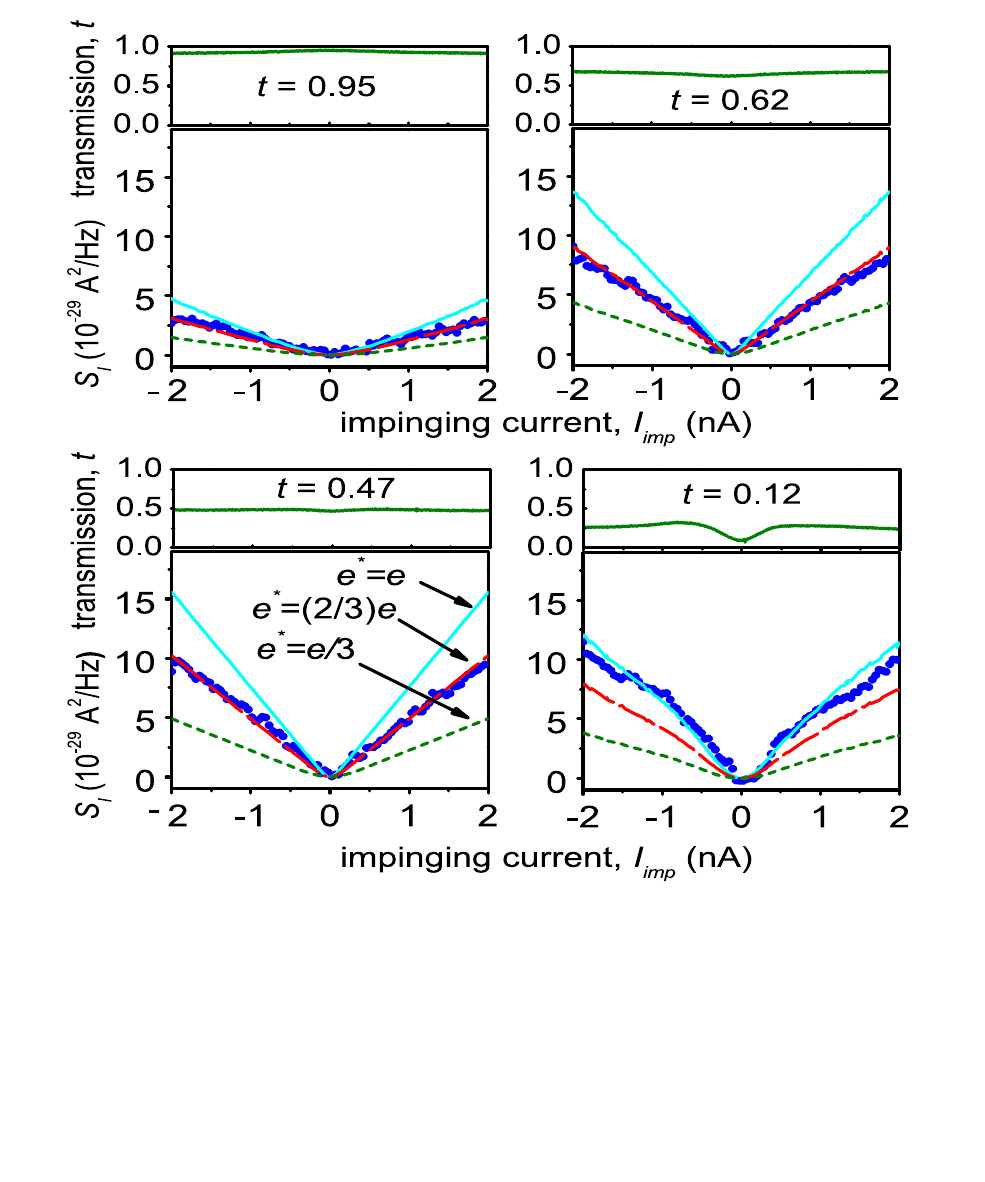}
\caption{Non-linear transmission (upper panels) and spectral density $S_I$ (lower panels) for few values of transmission \emph{t} of the constriction.  Blue dots are the measured data; cyan solid line is the expected spectral density for quasiparticle charge $e^*=e$; red dashed line for $e^*$=(2/3)$e$; olive dotted line for $e^*$=$e$/3.}
\label{fig:figure3}
\end{figure}
\begin{figure}[b]
\includegraphics[viewport=0.5in 1.4in 3.6in 3.35in,width=0.8\columnwidth]{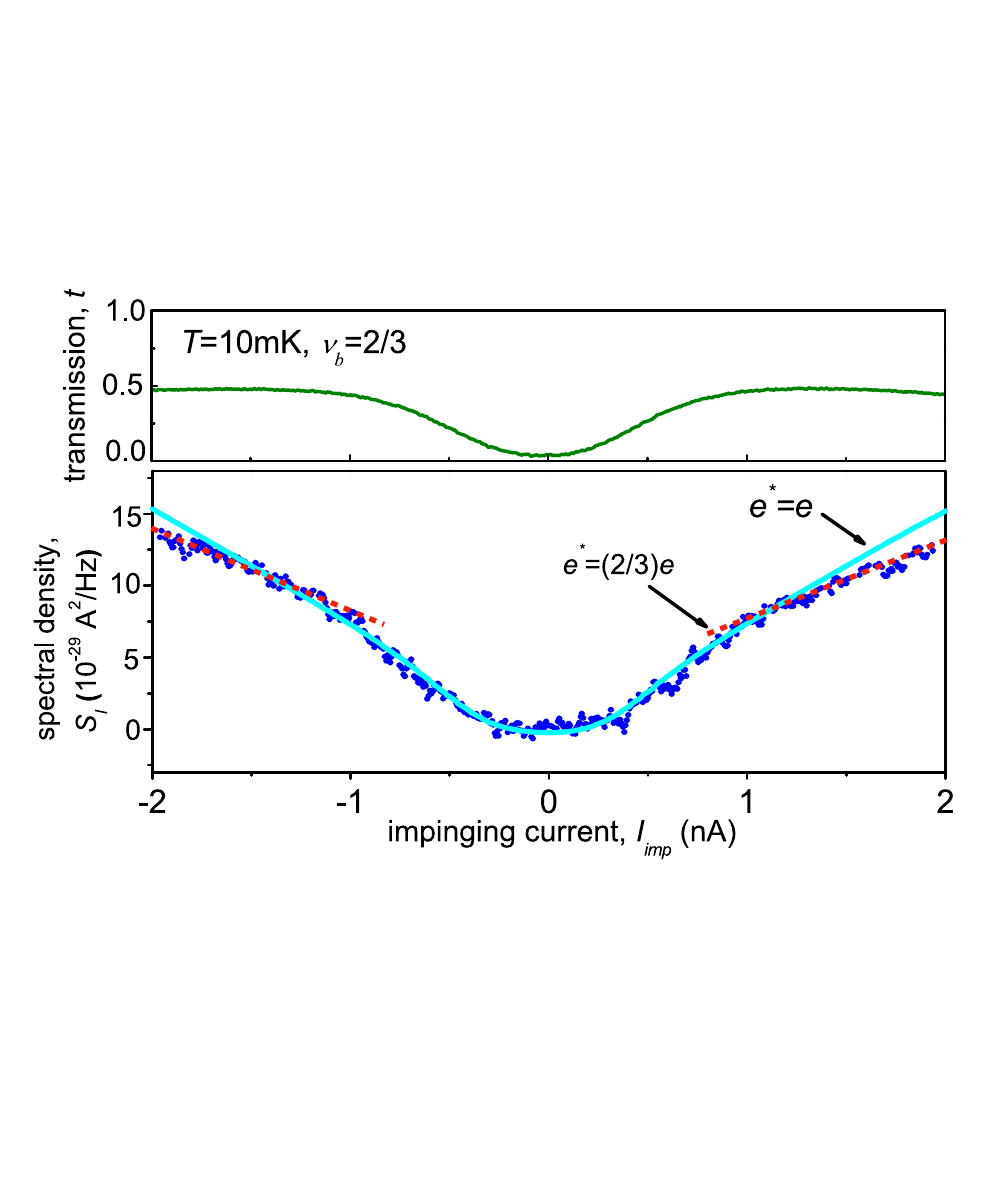}
\caption{At low transparency of the constriction and small $I_{imp}$, the tunneling quasiparticle becomes \emph{e} (linear \emph{t}=0.05).  At higher currents, the transmission of the constriction increases, leading to tunneling of quasiparticles with charge approximately $e^*$=(2/3)$e$.  Blue dots are the measured data; cyan solid line is the expected spectral density for quasiparticle charge $e^*=e$ and red dotted line for $e^*$=(2/3)$e$.}
\label{fig:figure4}
\end{figure}
Zero shot noise on conductance plateaux was observed before for numerous filling factors in the bulk [e.g., ~\cite{twenty, twentyfive}]; here we provide an example for $\nu_b$=2/5 and $\nu_c$=1/3 (fig.~\ref{fig:figure2}b) - with zero shot noise measured on the 1/3 plateau. However, surprisingly, this was not the case here for $\nu_b$=2/3 and $\nu_c$=1/3.  As shown in Fig. 2b, the measured shot noise is finite, suggesting a different picture of edge reconstruction in the $\nu$=2/3 case.  Since this behavior repeated itself in all the measured samples, we adopted the notion of a single chiral-composite-edge channel, with transmission $0\leq t \leq1$, determined by the split-gate voltage.  Note that this is inconsistent with the measurements of Saminadayar \emph{et al.} ~\cite{seven}, who measured a quasiparticle charge of \emph{e}/3 at $\nu_B$=2/3 assuming transport through two distinct channels.

The measured shot noise spectral density as a function of $I_{imp}$, at a few selected values of constriction's transmission, is shown in Fig.~\ref{fig:figure3}.  The best fits to the data using the measured differential transmission and electron temperature are shown.  The charge over the entire range of the transmission (from very close to unity to $\sim$0.3) fits excellently to a value $e^*$=(2/3)$e$.  Upon depleting the constriction further the differential transmission became highly current dependent and the $S_I$ vs $I_{imp}$ plot developed two distinct slopes: at the low range of current the quasiparticle charge was \emph{e} while at higher currents the charge dropped to approximately (2/3)\emph{e} (see Fig.~\ref{fig:figure4}; note that at the high current range the effect of temperature is insignificant and the charge can be estimated from the slope of the spectral density).  Indeed, the resultant electrons tunnelling was predicted and had been already observed in `simpler' FQHE states ~\cite{twentyfive}.  Since the transmission increases with current (expected in a chiral Luttinger liquid), the charge reverts back to that of a quasiparticle.


\begin{figure*}[t]
\includegraphics[viewport=3in 0.6in 9in 4.6in,width=0.75\columnwidth]{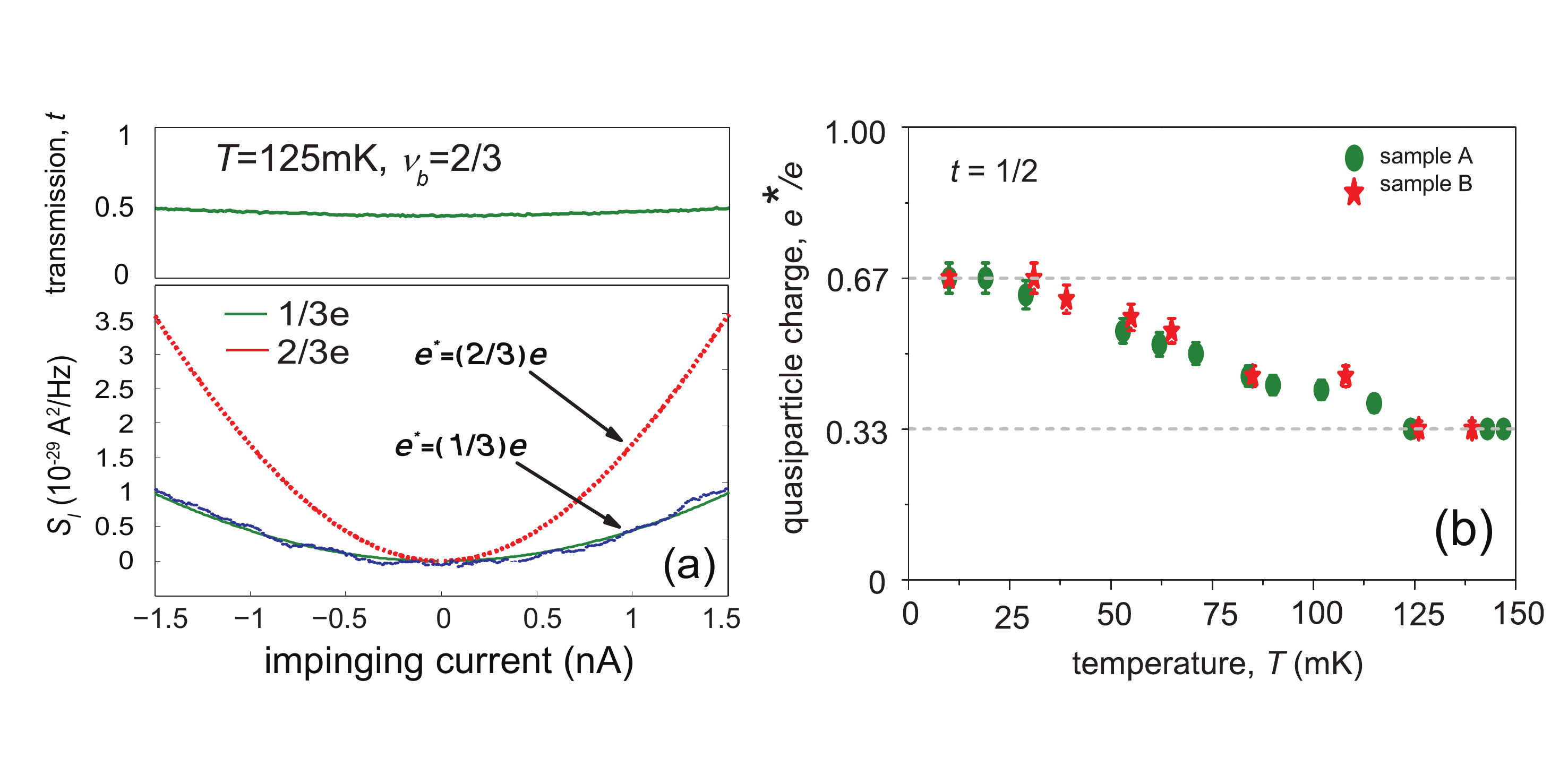}
\caption{Shot noise at elevated temperatures.  (a) Transmission and spectral density (blue dots) measured at \emph{T}=125mK and transmission \emph{t}$\sim$  1/2 - quasiparticle charge is $e^*$=$e$/3 (red dotted and olive solid lines are the expected spectral density for quasiparticle charge $e^*$=(2/3)$e$ and $e^*$=$e$/3 respectively) (b) Normalized quasiparticle charge as a function of electron temperature for the two samples fabricated on two different heterostructure (see text for details of the materials).}
\label{fig:figure5}
\end{figure*}


In Fig.~\ref{fig:figure5}a we show a plot of the spectral density measured at an electron temperature of 125mK, $\nu_b$=2/3, and \emph{t}$\sim$0.5.  The deduced quasiparticle charge is in good agreement with \emph{e}/3.  Note that the bulk longitudinal resistance $R_{xx}$ was still zero, verifying the sturdiness of the 2/3 state.  The temperature evolution of the quasiparticle charge at $t \sim 0.5$, measured with the two different heterostructures, is shown in Fig.~\ref{fig:figure5}b.  As the temperature increased the charge evolved smoothly from (2/3)\emph{e} to \emph{e}/3.  This trend repeated for $1\leq t\leq0.4$.  This observation, reminiscent of the behavior of the quasiparticle charge in the 2/5 and 3/7 states ~\cite{twentyone}, may suggest that quasiparticles of charge \emph{e}/3 carry the current in the 2/3 state, however, `bunching' takes place at the low temperature regime.

One of the important findings here is the absence of noise at \emph{t}=1/2.  Since the electron density gradually decreases toward the constriction edges, edge channels for the  $\nu$=2/3 are expected to comprise of two independent modes: an outer channel with conductance (1/3)$e^2/h$ and an inner one with conductance (2/3)$e^2/h$, which in turn is believed to be a composite channel - made of a charged mode and a counter-propagating neutral mode.  This picture, if true, would lead to zero shot noise at \emph{t}=1/2, with a visible conductance plateau.  However, the observed finite shot noise on the 1/3 plateau seems to imply that the picture is more complicated, and the edge reconstructs to a single composite mode with conductance (2/3)$e^2/h$.  However, the reason for the conductance plateau as function of slit-gate voltage is not clear.  Another important observation is the evolution of the backscattered charge from $e^*$=(2/3)\emph{e} to $e^*$=\emph{e}/3 as the temperature increased to about 125mK.  Indeed, at finite temperatures the neutral mode is expected to decay over a finite length scale leaving behind only the charged mode [of conductance (2/3)$e^2/h$], with quasiparticle charge $e^*=e$/3 that is expected from the theoretical predictions.

We acknowledge the partial support of the Israeli Science Foundation (ISF), the Minerva foundation, the German Israeli Foundation (GIF), the German Israeli Project Cooperation (DIP), the European Research Council under the European Community's Seventh Framework Program (FP7/2007-2013)/ERC Grant agreement No. 227716, and the US-Israel Bi-National Science Foundation.  N.O. acknowledges support from the Israeli Ministry of Science and Technology.


\begin{thebibliography}{30}
\bibitem{twentyfour} Sample's equilibrium thermal noise ($\sim 2 \times 10^{-29}$ A$^2$/Hz) and the cryogenic preamplifier's current noise ($2 \times 10^{-28}$ A$^2$/Hz) sum up to a background level of the order $\sim 2 \times 10^{-28}$ A$^2$/Hz, while shot noise values are in the $10^{-30}$ A$^2$/Hz range.
\end{thebibliography}

\end{document}